# The adjustable thermal resistor by reversibly folding a graphene sheet

Qichen Song[1,2, #†], Meng An[1,2, #], Xiandong Chen[1,2], Zhan Peng[1,2],

Jianfeng Zang[3,4, *], Nuo Yang[1, 2, *]

[1] State Key Laboratory of Coal Combustion, Huazhong University of Science and Technology, Wuhan 430074, People's Republic of China

[2] Nano Interface Center for Energy (NICE), School of Energy and Power Engineering, Huazhong University of Science and Technology (HUST), Wuhan 430074, People's Republic of China

[3] School of Optical and Electronic Information, Huazhong University of Science and Technology, Wuhan 430074, People's Republic of China

[4] Innovation Institute, Huazhong University of Science and Technology, Wuhan 430074, People's Republic of China

# Q.S. and M.A. contributed equally to this work.

† Current address: Department of Mechanical Engineering, Massachusetts Institute of Technology, 77 Massachusetts Avenue, Cambridge, MA 02139, USA.

* To whom correspondence should be addressed. E-mail: (N.Y.) nuo@hust.edu.cn (J. Z.) jfzang@hust.edu.cn

**ABSTRACT**

**Phononic (thermal) devices such as thermal diode, thermal transistors, thermal logic gates, and thermal memories have been studied intensively. However, tunable thermal resistors have not been demonstrated yet. Here, we propose an instantaneously adjustable thermal resistor based on folded graphene. Through theoretical analysis and molecular dynamics simulations, we study the phonon-folding scattering effect and the dependence of thermal resistivity on the length between two folds and the overall length. Further, we discuss the possibility to realize the instantaneously adjustable thermal resistor in experiment. Our studies bring new insights in designing thermal resistor and understanding thermal modulation of 2D materials by adjusting its basic structure parameters.**

**KEYWORD:** Adjustable thermal resistor, folded graphene, size effect, phonon transport, phonon scattering

**Introduction**

A detailed understanding of phonon transport enables people to manipulate heat flow at nanoscale and design the phononic devices[1] based on electronic analogs, e.g. thermal diode[2], thermal transistor[3], thermal logic gates[4], etc. In electronic circuits, a variable electrical resistor (potentiometer) tuning dynamically the electric load is essential to multiple applications. However, the thermal counterpart to instantaneously adjustable electrical resistor is rather interesting but has not yet been proposed.

The basic requirement of an adjustable electrical resistor is a simple resistance-position relationship, for example, linear characteristics, i.e. the electrical resistance has a linear dependence on the distance between positions of two contacts. To develop a thermal version of the linear characteristics, it is expected that the thermal resistance of candidate material should have a linear dependence on parameters that can be easily controlled. Recently, scientists demonstrates a great potential in tuning transport properties of graphene simply by controlling the deformation status[5], creating even more probabilities beyond its high electrical[6] and thermal[7] conductivities. For nano devices based on graphene, it is viable to change its shape (e.g. folding) due to its high flexibility[8]. Recent reports show that the graphene nanoribbon with folds can modulate electron transport[9,10], phonon transport[11,12] and mechanical properties[13]. Particularly, its thermal conductivity can be modulated with different number of folds[14]. As is well known, the size effect is an important issue in nanostructures[12]. However, most of previous studies focused on the graphene nanoribbon of a finite size. More recently, the reversibility of folding and unfolding of large-area graphene sheets has been demonstrated in experiment[5], making it possible to generate certain numbers of folds in an initially planar graphene or eliminate the folds from a folded graphene.

Here, we propose an instantaneously adjustable thermal resistor based on the folding effects on large-area graphene. We study folded graphene with various shape parameters. Moreover, we introduce a new phonon scattering regime, named as phonon-folding scattering, which stems from this novel folded structure and explain the length-dependent behavior of thermal resistivity of folded graphene.

**Theoretical Model**

Experimentally, the folding and unfolding of graphene sheet can be controlled by the substrate deformation instantaneously[5]. That is, the number of folds and the degree of folding can be controlled by changing the strain of the substrate, the shear modulus of the substrate, and the adhesion energy between the graphene and the substrate. The schematic of adjustable thermal resistor based on graphene with folds is illustrated in Fig. 1(a). The device, firstly, takes the advantage of the fact that phonon mean-free path (MFP) is relatively large in suspended graphene (100 nm ~ 600 nm)[11,15] and phonon transport in graphene can be affected by its structure. Secondly, the resistance depends on both the number of folds and the distance between two folds due to folding. The device can achieve linear characteristics of thermal resistance and the mechanism can be explained by the following theoretical model.

It is known as the Casimir limit[16] that the phonon MFP and thermal conductivity are limited by the system size. The effective MFP, denoted as $l_{eff}$, in a finite system is given by Schelling[17] *et al.* as,

$$\frac{1}{l_{eff}} = \frac{1}{l_\infty} + \frac{2}{L},  \qquad (1)$$

where $l_\infty$ is the phonon MFP for the infinite system and $L$ is the length of finite system. Eq. (1) is known as finite-size effect due to the fact that phonon transport is affected by anharmonic phonon-phonon scattering and scattering with heat source and sink. Between heat source and heat sink, some phonons travel ballistically across the system and the MFP of those phonons is limited by system size $L$. For those phonons, the average distance they travel is $L/2$. We know that there is a linear relationship between thermal conductivity and MFP as $\kappa = C_v v_g l$. When the size effect of specific heat $C_v$ and group velocity $v_g$ is negligible, it can be derived from Eq. (1) that the resistivity $r$ has a linear dependence on $1/L$, as

$$r = \frac{1}{\kappa} = \frac{1}{C_v v_g}\left(\frac{1}{l_\infty} + \frac{2}{L}\right). \qquad (2)$$

In nanowires or graphene ribbons, $L$ is the longitudinal length[18].

When there are more than one factor limiting the phonon MFP simultaneously,

according to the Matthiessen rule[19], the total scattering event can be described as follows,

$$\tau_{scatt}^{-1} = \sum_j \tau_{scatt,j}^{-1}, \tag{3}$$

where $\tau_j$ is the relaxation time of the scattering process $j$. When Eq. (3) is multiplied by $1/v_g$, then the inverse effective MFP can be written as $l_{eff}^{-1} = \sum_j l_j^{-1}$, where $l_j$ is the characteristic length of the scattering process $j$.

In a folded graphene, the scattering processes include intrinsic anharmonic phonon-phonon scattering, finite-size effects, and phonon-folding scattering meaning that phonons can be scattered by folds. As shown in Fig. 1(b), between two folds the distance across plane is denoted by $L_{plane}$. For those phonons that move ballistically in plane, the term in the inverse effective MFP corresponding to phonon-folding scattering is $2/L_{plane}$. Nevertheless, phonon-folding scattering is different from finite-size effects as some phonons can pass through the folds without being scattered. For instance, it is entirely possible that a vibration mode along y direction is not affected by the folds at all. So the scattering term $2/L_{plane}$ overestimates the strength of phonon-folding scattering. Here, we add a coefficient $\alpha$ in front of $2/L_{plane}$ to include this effect. It is noted that phonon-folding scattering is similar to impurity scattering, grain-boundary scattering, and isotope scattering, etc., since they are all disturbance to perfect crystal and are usually be assumed to be independent process[20,21]. Therefore, the phonon-folding scattering can also be treated as independent process with finite-size effects and intrinsic anharmonicity and the thermal resistivity of folded graphene is:

$$r = \frac{1}{\kappa} = \frac{1}{C_v v_g} \left[ \frac{1}{l_\infty} + 2 \times \left( \alpha \frac{1}{L_{plane}} + \frac{1}{L} \right) \right]. \tag{4}$$

As $\kappa_\infty = C_v v_g l_\infty$, then we have

$$r = \frac{1}{\kappa} = \frac{1}{\kappa_\infty} \left[ 1 + 2 l_\infty \left( \frac{\alpha}{L_{plane}} + \frac{1}{L} \right) \right]. \tag{5}$$

Eq. (5) is the main theoretical model and it will be proved by MD simulation results.

To design the adjustable thermal resistor, there are two possible strategies. One way is to change the length of structure and the other one is to tune its thermal resistivity directly. The former approach is hard to achieve at nanoscale since the crystal growth usually takes a long time and the size of graphene will not be changed after being fabricated. So we move our attention to the latter approach by changing thermal resistivity while keeping $L$ invariant. Now, through modulating the distance between two folds, $L_{plane}$, i.e. changing the number of folds, n, we can adjust thermal resistance of one single piece of graphene.

**MD Method**

In MD simulations, the simulation cell of the folded graphene is built as shown in Fig. 1(b-c). The length of graphene is characterized by $L = (n+1) \cdot L_{plane} + n \cdot L_{fold}$, where n is the number of folds and $L_{plane}$ and $L_{fold}$ are the length of flat part and folded part, respectively. The width of the simulation cell is set as 2.13 nm (10 atoms in each layer) for the reason that the thermal conductivity of cases with larger width is independent on width. Periodic boundary condition is applied in the y direction, and atoms in the two ends are fixed. By the Nosé-Hoover thermostat method[22,23], five nearest layer atoms to one fixed boundary are maintained at $T_h = (1+\Delta) \cdot T$, while those nearest to the other side are at $T_c = (1-\Delta) \cdot T$, where $T = 300K$ and $\Delta = 0.1$. The substrate effect is included to imitate the experiment settings, where the couplings are described by the Lennard-Jones 12-6 potential[24].

A Morse bond and a harmonic cosine potential energy including two-body and three-body potential terms[25-27] are used to describe the bonding interaction between carbon atoms. Although the force field potential is developed by fitting experimental parameters for graphite, it has been testified by the calculation of thermal conductivity of graphene[14,28] and carbon nanotubes[25]. The simulation domain is bounded with two Lennard-Jones (LJ) 12-6 potential[24] walls in the z direction that enclose atoms of the top plane and bottom plane. In all MD simulations, the Velocity-Verlet algorithm[29] is used to integrate the differential equations of motions. The time step of 0.5 fs is adopted and the total simulation time is set as 3 ns. At the very beginning of the simulation, the distance between nearest plane layers is 0.474 nm. For the first $3\times10^5$ steps, the positions of the both substrates are moved towards each other at a small enough pace (~$10^{-6}$ nm per step), small enough compared with the movement of the atom at each step (~$10^{-4}$ nm). Then during the rest of the evolution, the distance between nearest plane layers is relaxed under the VDW force. After relaxation, the inter-plane distance approaches the same value, 0.350 nm, and the fold length also approaches the same values 0.737 nm. Note that the fold length decreases as the inter-plane distance decreases. Our previous work[14] has observed that the thermal resistivity increases with the decreasing inter-plane distance due to the fact that compressing inter-plane distance could enhance phonon-phonon scattering. In addition, literature shows that as the curvature for the folds increases, the thermal resistivity increases[30]. In this work, we are looking for the regime that resistivity solely depends on the characteristic length, thus we use the same inter-plane distance, same fold length and same curvature for all calculation cases.

The total heat flux ($J_t$) is recorded by the average of the input/output power at the two baths as

$$J_t = \frac{1}{N_{T_h(T_c)}} \sum_{i=1}^{N_{T_h(T_c)}} \frac{\Delta \varepsilon_i}{2\Delta t}, \tag{6}$$

where $\Delta \varepsilon$ is the energy added to/removed from each heat bath ($T_h$ or $T_c$) at each step $\Delta t$. The total heat flux can be divided into the in-plane flux ($J_i$) and the inter-plane heat flux ($J_{int}$), where $J_t = J_i + J_{int}$. In order to obtain $J_i$, we record the heat flux carried by inter-plane interaction, the van der Waals (VDW) force, based on[31,32]

$$J_{int} = \sum_{i \in A, j \in B} J_{i \to j} = \sum_{i \in A, j \in B} \frac{1}{2} \langle \mathbf{F}_{ji} \cdot \mathbf{v}_j + \mathbf{F}_{ji} \cdot \mathbf{v}_i \rangle, \tag{7}$$

where $\mathbf{F}_{ji}$ is the VDW force on atom $j$ by atom $i$, and A and B represent the two group of atoms separated by a cross section that the heat energy passes through. With that, the in-plane heat flux can be calculated by

$$J_i = J_t - J_{int}. \tag{8}$$

Here we are more interested in the in-plane phonon transport. Based on the format of Fourier's Law, the thermal conductivity can be calculated by

$$\kappa_i = -\frac{J_i}{A\nabla T}, \tag{9}$$

where $J_i$ is the heat current along the structure, and $A$ is the cross section area and $\nabla T = (T_h - T_c)/L$. The results presented here are the average of 12 independent simulation cases with different initial conditions and the error bar of thermal conductivity is the deviations of 12 simulation results. In addition to the folded structures, the thermal conductivity of planar graphene is also calculated for comparison.

**Results and Discussions**

Firstly, we studied both a planar graphene and a folded graphene of which the length $L$ are kept the same as 36.7 nm. The temperature distribution profiles of the folded graphene and the planar graphene are presented in Fig. 1(d). The temperature in folded graphene shows a stair-step mode, which is obviously different with that in planar graphene. Each sharp change point in temperature profile indicates a larger thermal resistivity at the folds than in the plane. In addition, the linear characteristic of the thermal resistivity $r$ is simulated when we keep the overall length $L$ to be 36.7nm and change the distance between two folds, $L_{plane}$. That is, the number of folds is tuned from 1 to 7.

As shown in Fig. 2(a), it is found that $r$ is dependent on $1/L_{plane}$ in structure. It reveals that the phonons are scattered at the folds and the MFP is restricted by $L_{plane}$. More interestingly, the thermal resistivity $r$ of folded graphene depends linearly on $1/L_{plane}$ for all calculation cases, which proves that our theoretical model of Eq. (5) is valid and reasonable.

Moreover, we studied the structure of which the $L_{plane}$ are kept the same, while the number of folds (n) and lengths ($L$) are different. The systems of different $L$ are explored in order to find the influence of initial length of graphene before folding. As $L_{plane}$ is fixed, the enlargement of $L$ is solely related to increase of the number of folds. The resistivity $r$ with respect to $1/L$ is shown in Fig. 2(b). When the $L$ becomes larger, $r$ increases as $r \sim 1/L$, similar to trend of planar graphene. This is because the decreased length $L$ confines phonon modes that exist in the structure. Meanwhile, the linear relationship confirms that our theoretical model of Eq. (5) is valid. When we extrapolate the linear fitting curve to $1/L = 0$, it is found that the thermal resistivity of infinitely large folded graphene is 4.4 times as high as the value of graphene.

Admittedly, precisely controlling the inter-plane distance in the process of folding and unfolding of large-area graphene, especially in nanoscale, is challenging and subject to large uncertainties. Nevertheless, the heat energy cannot be transferred via the weak VDW forces when the inter-plane distance reaches beyond the radius of VDW forces. Therefore, we proceeded to compare the in-plane heat flux to the total heat flux, as shown in Fig. 3. As $L_{plane}$ decreases with the increased number of folds, the inter-plane flux is reduced since more folds leads to a smaller plane area to transport heat energy by VDW forces. Moreover, the ratio of inter-plane heat flux to the total heat flux tends toward the convergent value around 20%, with the increasing number of folds. In order to mimic experimental conditions (the inter-plane flux is usually negligible), the inter-plane heat flux is subtracted from the total heat flux.

The size dependence of revised resistivity on the characteristic sizes, $1/L$ and $1/L_{plane}$, are plotted in Fig. 4(a-b). It is shown that the thermal resistivity is still a linear function of the characteristic size, $1/L$ and $1/L_{plane}$. Note that in Fig. 4(b), we extrapolate the thermal resistivity to $L \rightarrow \infty$. If we assume that the thermal resistivity in bulk folded graphene and bulk planar graphene are approximately equivalent, the ratio of their thermal resistivity is,

$$\eta = \left.\frac{r_{folded}}{r_{planar}}\right|_{\frac{1}{L}=0} = 1 + \frac{2\alpha l_\infty}{L_{plane}}. \tag{10}$$

The phonon MFP in folded graphene can be calculated using the slope of thermal resistivity curve in Figure 4(b). Therefore, the coefficient $\alpha$ is,

$$\alpha = \frac{(\eta-1)L_{plane}r_{\infty}}{\frac{\Delta r}{\Delta(1/L)}}. \tag{11}$$

For folded graphene of $L_{plane}$ = 6.75 nm, $\eta$ = 2.41, $\alpha$ = 0.20, which is obtained from Figure 4(b). Now, we can compare the scattering term of finite-size effects and phonon-folding scattering. The ratio of relaxation time of the two type of scattering process is $\frac{\tau_{finite}}{\tau_{folding}} = \frac{L_{plane}}{\alpha L}$. When $L$ = 33.75 nm, the ratio is 1 meaning that finite-size effects and folding scattering have similar scattering rates. When $L$ is much larger than 33.75 nm, the finite-size effects dominate and folding scattering becomes negligible. When $L$ is smaller than 33.75 nm, the finite-size effects are less significant than folding scattering.

We believe that the phonon-folding scattering regime is indeed phonon scattering at the entire region of fold induced by the folding structure. To further understand phonon-folding scattering, phonon power spectrum of atoms in the plane and at the fold at room temperature is calculated. As shown in Fig. 5, the spectrum describes the power carried by phonon per unit frequency. In the MD simulations to record the spectrum, there is no external thermostat applied.

A higher value of phonon spectral density (PSD) at a certain frequency $f$ indicates that there are more phonons occupying states, whereas a zero value of PSD means that there are no such phonons existing. As is illustrated in Fig. 5, the power spectra of atoms in the plane are similar to graphene, where the in-plane/out-of-plane mode phonons are mainly distributed in high-frequency/low-frequency range[33,34]. However, we do not observe the same scenario for atoms in the power spectra of atoms at the fold. Its spectra have significant differences from those of atoms in the plane because the cylindrical distribution of atoms leads to a mix of vibrations along different directions. The spectra at the fold are similar to those of nanotube, of which the highest peak locates within the high-frequency range[35,36]. The mixed vibrations along different directions is also observed in the folded graphene nanoribbon[14].

When phonons pass the fold, some phonons along z direction need to change from the out-of-plane into mixed mode, and change back to out-of-plane mode after passing the folds. There are appreciable differences between the z-direction PSD of atoms at the fold and in the plane in low-frequency range (0 ~ 10 THz). The two peaks in power spectrum of atoms in the plane disappear when it comes to the fold, which means that low-frequency phonons are depressed. These mismatch behaviors of phonons will induce phonon scattering to redistribute the phonon energy to allow phonons to pass through the fold.

Finally, we would like to discuss the possibility of the experimental realization of the instantaneously adjustable thermal resistor. Our previous report[5] has shown that the number of folds/the length between two folds of large-area graphene can be easily controlled by changing the strain of the substrate, the shear modulus of the substrate, and the adhesion energy between the graphene and the substrate. This enables us to tune $L_{plane}$ instantaneously while keeping $L$ as a constant. Meanwhile, in this work, we have found that thermal resistivity calculated with/without subtracting inter-plane interactions both depend linearly on $1/L_{plane}$. This ensures linear characteristics of thermal resistance. Based on these discussions, we believe that the instantaneously adjustable thermal resistor could be realized in the near future.

**Conclusion**

Similar to other nanostructures, the overall-length dependence of the thermal resistivity in folded graphene arises from the finite-size effects. Interestingly, the resistivity depends linearly on the length between two folds. The underlying physical mechanism is phonon-folding scattering, i.e. phonon scattering due to mode mismatch between plane and fold, which has great impact on those phonons with MFP longer than the length between two folds.

Our results are of vital importance for building instantaneously adjustable thermal resistor. Since the number of folds/the length between two folds can be easily controlled by the substrate deformation, the thermal resistance is directly determined by the strain executed on the substrate. Besides graphene, we believe that other 2D materials with large phonon MFP can also be applied as an adjustable thermal resistor by folding due to the applicability of phonon-folding scattering effect. The realization of the adjustable thermal resistor not only completes the spectrum of thermal analogs of electronics but leads to a brighter future of more possibilities for thermal devices based on adjustable thermal resistor.

**Acknowledgement**

N. Y. was supported by the National Natural Science Foundation of China (51576067). J. Z. thanks the support from the National Natural Science Foundation of China (51572096) and the National 1000 Talents Program of China tenable in Huazhong University of Science and Technology (HUST), China. We are grateful to Zelin Jin, Yingying Zhang, Dengke Ma and Shiqian Hu for useful discussions. The authors thank the National Supercomputing Center in Tianjin (NSCC-TJ) for providing assistance in computations.

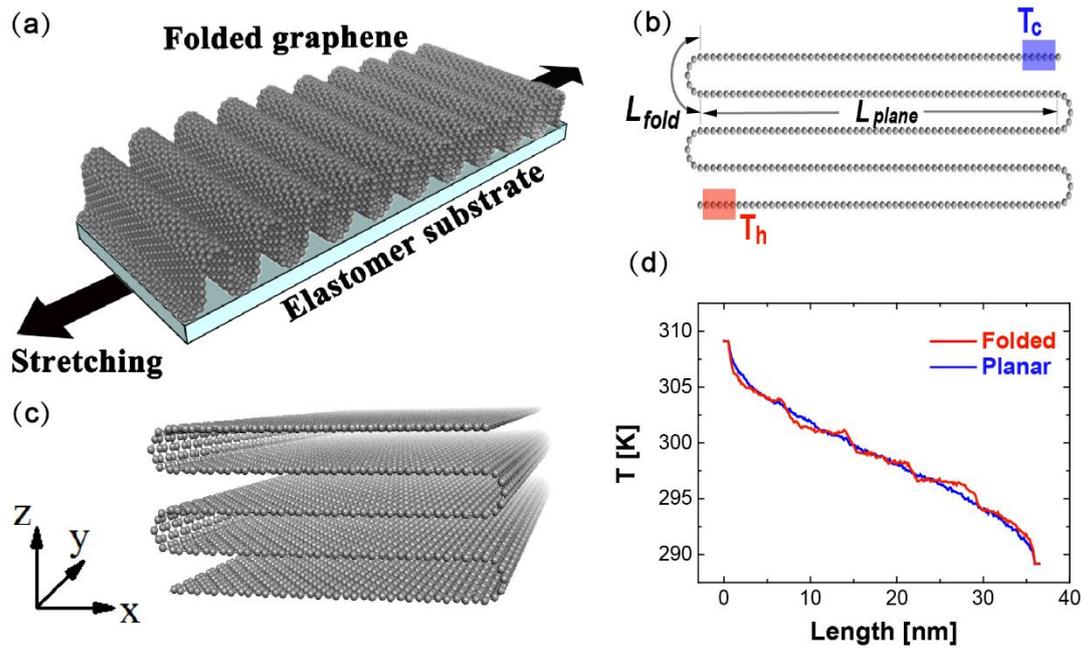

Figure 1. (a) Schematic illustration of an instantaneously adjustable thermal resistor. (b) The side view of the structure of folded graphene before relaxation. $L_{plane}$ and $L_{fold}$ are 6.75 nm and 0.737 nm, respectively. The number of folds is 4 and the length $L$ of the structure is 36.7 nm. The initial inter-plane distance before relaxation is 0.474 nm. Nosé-Hoover thermostat with high temperature ($T_h$) and low temperature ($T_c$) is applied to the bottom (red) and top (blue) graphene sheets, respectively. (c) A projection view of the structure before relaxation. The width of the system can be treated as infinitely large with periodic boundary condition along y direction. (d) The stair-step temperature distribution profiles in the structure along the longitudinal direction, compared with that in planar graphene of same length.

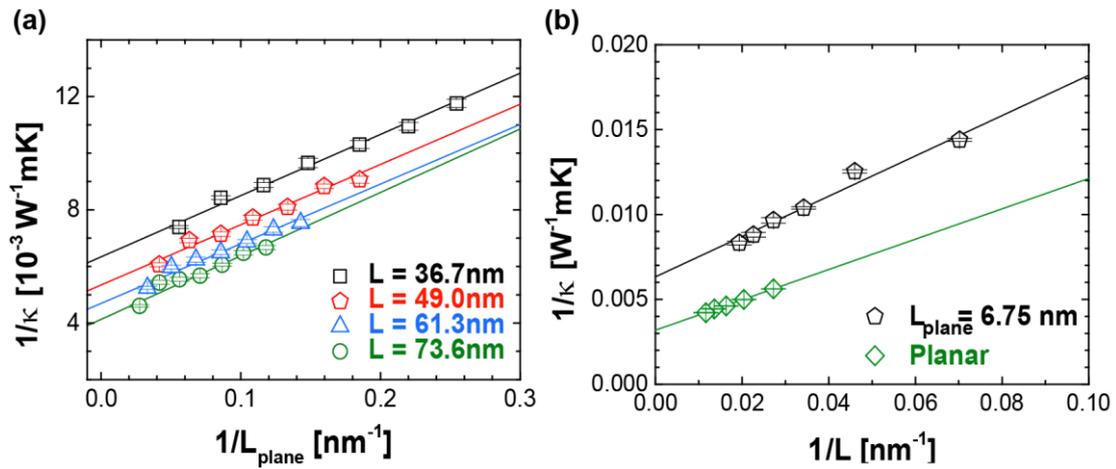

Figure 2. (Color online) (a) The size dependence of the resistivity of folded graphene on the reciprocal of length between two folds, $1/L_{plane}$. The fitting curves are based on Eq. (5). Note that the length of the folded graphene is fixed, which means $L_{plane}$ decreases as the number of folds (n) increases. The number of folds, n, ranges from 1 to 7, corresponding to seven data points in the same color from left to right. (b) The size dependence of the resistivity of planar/folded graphene on the reciprocal of length, $1/L$. The fitting curves are based on Eq. (5). Note that the distance across plane between two folds, $L_{plane}$, is constant in folded graphene, which means that length, $L$, varies with number of folds (n). The number of folds, n, ranges from 2 to 7, corresponding to the six data points in black from right to left.

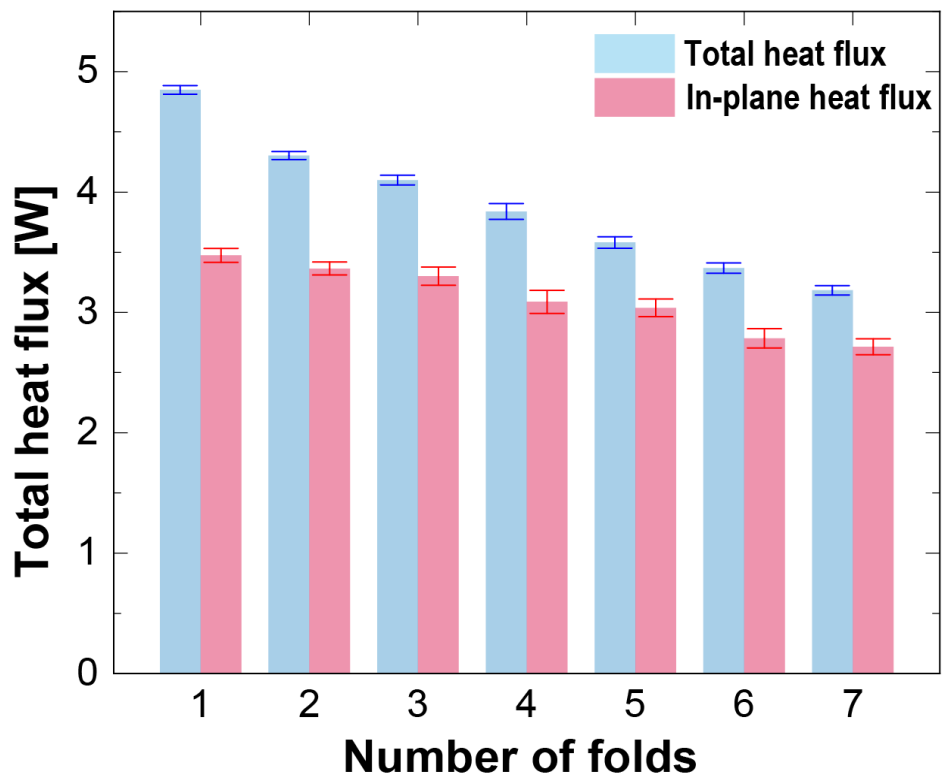

Figure 3. (Color online) Histogram of the total heat flux $J_t$ calculated by Eq. (6) and the in-plane heat flux $J_i$ calculated by Eq. (7).

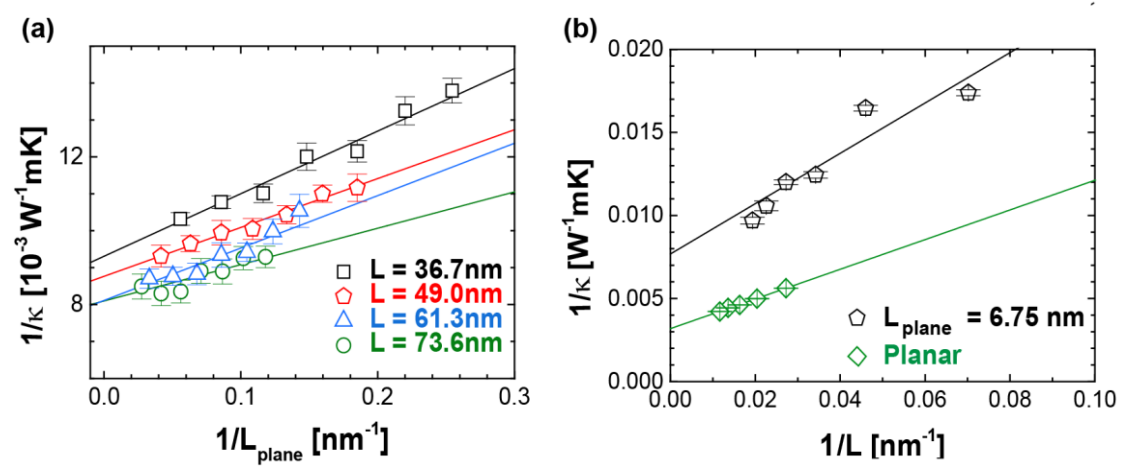

Figure 4. (Color online) (a) The dependence of revised resistivity of folded graphene on $1/L_{plane}$. (b) The dependence of revised resistivity on $1/L$ of folded and planar graphene.

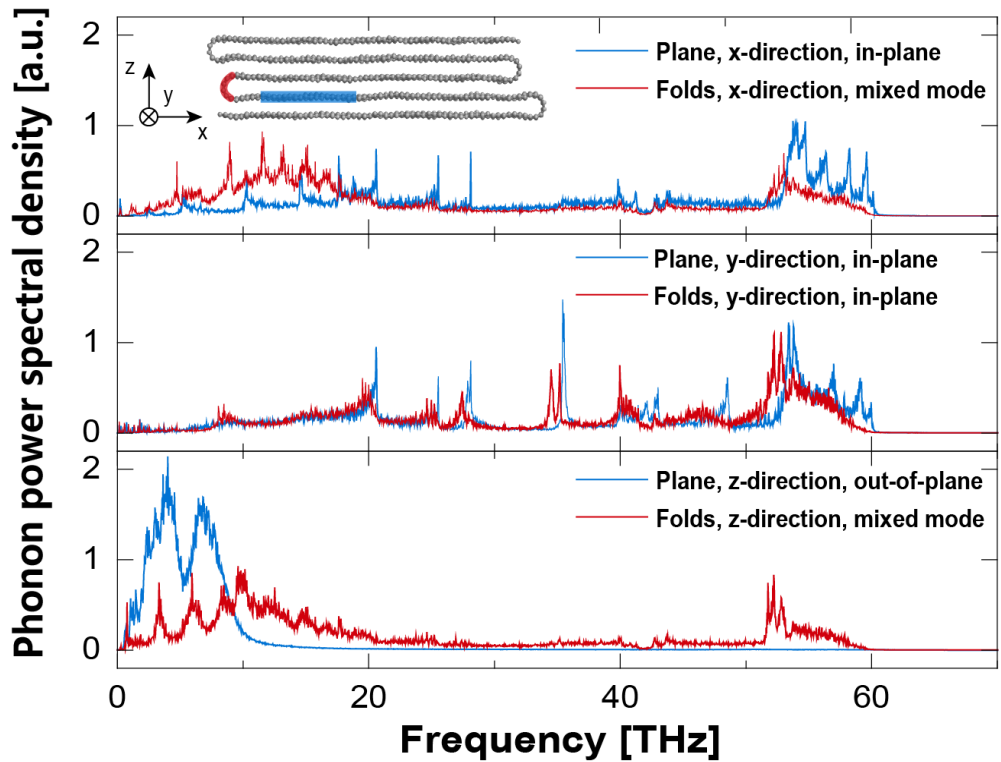

Figure 5. (Color online) The phonon power spectral density along different directions of atoms in the plane and in the fold structure, obtained by calculating the Fourier transform of recorded velocities along corresponding directions of selected atoms. The details of the structure for calculation can be found in the caption of Fig. 1. Velocities of all atoms in the red area (60 atoms) are recorded and 60 atoms evenly distributed in the blue area are selected to represent the vibrational properties of plane. Note that in the plane region, x-direction and y-direction mode are in-plane mode and z-direction mode is out-of-plane mode. In fold region, only y-direction mode is in-plane while x-direction and z-direction mode are mixed mode, which contains both in-plane and out-of-plane modes.